\documentclass[aps,pra,twocolumn,amsmath,amssymb,showpacs,fixfloats]{revtex4}
\newcommand{\be}{\begin{equation}}
\newcommand{\ee}{\end{equation}}
\newcommand{\bea}{\begin{eqnarray}}
\newcommand{\eea}{\end{eqnarray}}
\usepackage[dvips]{epsfig}
\usepackage{graphicx}
\usepackage[T1]{fontenc}

\begin{document}

\title{Feshbach resonances in ultracold atom-molecule collisions}
\author{Andrea Simoni and Jean-Michel Launay}
\affiliation{Institut de Physique de Rennes, UMR 6251 du CNRS and Universit\'e de
Rennes 1, 35042 Rennes Cedex, France}
\author{Pavel Sold\'{a}n}
\affiliation{Department of Chemical Physics and Optics, Faculty of Mathematics and Physics,
Charles University in Prague, Ke Karlovu 3, 121 16 Prague 2, Czech Republic}

\date{\today}


\begin{abstract}

We investigate the presence of Feshbach resonances in ultracold
alkali-dialkali reactive collisions. Quantum scattering calculations are performed on
a new Na$_{3}$ quartet potential energy surface. 
An analysis of scattering features is performed through a systematic variation
of the nonadditive three-body interaction potential.
Our results should provide useful information
for interpreting future atom-molecule collision experiments.

\end{abstract}

\maketitle


\section{Introduction}
In recent years, there has been a growing interest in
ultracold molecules~\cite{Kre08}, particularly in the production
and properties of the  molecules formed from ultracold atomic
gases~\cite{Hut06}. Photoassociation~\cite{Jon06} and Feshbach resonance
tuning~\cite{Koh06} are two main experimental techniques for a coherent
production of ultracold molecules from ultracold alkali-metal atoms.
In 2003 long-lived molecular Bose-Einstein condensates were created
from weakly bound homonuclear lithium and potassium dimers by
exploiting magnetically tunable Feshbach resonances between fermionic
isotopes~\cite{Grimm03a,Ket03a,Jin03a}.

While Feshbach resonances are always located at the highest
vibrational manifold of the dimer, photoassociation could in principle
allow access to low vibrational dimer states. In 2005 RbCs molecules were
created in their ground vibronic state~\cite{yale}. Very recently, several
different photoassociation schemes for molecular formation in the ground
vibronic state have also been developed~\cite{orsay,innsbruck1,freiburg,innsbruck2,jila1}.

For the correct interpretation of the forthcoming experiments
with cold molecular samples it is essential to understand
the atom-molecule and molecule-molecule interactions at sub-K
temperatures~\cite{Hut07}. Theoretical results have already been
published for the homonuclear X+X$_{2}$ ultra-low-energy collisions
with X\,=\,Li,\,Na,\,K~\cite{Sol02,Que04,Cvi05a,Que05,Que07,Cvi07}.
Isotopically heteronuclear Li+Li$_{2}$ ultra-low-energy collisions have
also been studied theoretically~\cite{Cvi07,Cvi05b,Li08a,Li08b,Li08c}.
Collision cross sections have been measured
experimentally for the Cs+Cs$_{2}$ ultracold inelastic
processes~\cite{pillet,weidenmuller}. Very recently reactive and
inelastic rate constants were measured for Li+Li$_{2}^{*}$ at room
temperature~\cite{Cop08}.

Cold collisions are known to be very sensitive to potential energy surfaces~\cite{Que04,Cvi07}, and
therefore experimental information is needed to improve
the corresponding theoretical models. In particular, knowledge of
low-energy resonance patterns often allows different properties of the
interaction potential to be determined with high accuracy. Such resonances
have been studied in great detail both theoretically and experimentally
in ultracold atomic gases; see {\it e.g.}~\cite{cs2,rb2,krb}.

On the other hand, very little is known about atom-molecule resonances in
ultracold collisions. Similarly to atomic
scattering the hyperfine-induced resonances could in principle exist at very
low collision energies. However, model calculations have shown that for
a general polarization they will be quenched by inelastic spin-exchange
transitions forming singlet molecules~\cite{lasphys}.

Alkali-metal dimers on the lowest electronic triplet manifold are only stable if they are in a
doubly spin-polarized state under collisions with doubly spin polarized
atoms (assuming that the relativistic spin interactions are neglected).
Unfortunately, for this specific polarization hyperfine-induced
resonances are prevented by symmetry (resonances induced by relativistic spin interactions are still possible).
In spite of this, long-lived three-atom complexes can in principle exist and give rise
to resonances. Such reactive resonances have been identified in reactive
collisions at room temperature~\cite{skodje}.

In this work, we focus on collisions of Na$_{2}$
molecules in the triplet electronic state with ground-state Na atoms. For this study
a new potential energy surface of Na$_{3}(1^{4}A_{2}')$ has been constructed.
The occurrence of long-lived three-atom resonances in such collision complex
is demonstrated in the ultracold regime for the first time.
We also study the dependence of collision cross sections on the potential
energy surface and we show that at least knowledge of two terms in the
cross section partial wave expansion is needed in order to characterize further the
three-body potential. 

\section{Potential energy surface}

\textit{Ab initio} calculations were performed using a single-reference
restricted open-shell variant~\cite{KHW93} of the coupled-cluster method
\cite{Cizek} with single, double and non-iterative triple excitations
[RCCSD(T)]. A basis set consisting of [12s,12p,5d,2f,1g]  basis functions
\cite{Nabas} was used for the dimer calculations, and the same basis
set without the g functions was used for the trimer calculations.
Electrons from the 1s orbital  on each sodium atom were not correlated in
the coupled-cluster calculations.  The three-atom interaction potential
was decomposed into a sum of pair-wise additive and non-additive contributions
\begin{equation}
\label{Eq1} V_{\rm trimer}(r_{12},r_{23},r_{13}) = \sum_{i<j}
V_{\rm dimer}(r_{ij}) + V_{3}(r_{12},r_{23},r_{13}).
\end{equation}
It has been shown by several authors that in the case of alkali-metal trimers the non-additive term $V_{3}(r_{12},r_{23},r_{13})$ is rather large and cannot be neglected \cite{Hig00,Sol03,Klo08}.
Interaction energies were calculated with respect to the separated-atoms
dissociation limit, and the full counterpoise correction of Boys and
Bernardi~\cite{BSSE} was employed to compensate for the basis set
superposition error in both the dimer and trimer calculations. All the
\textit{ab initio} calculations were performed using the MOLPRO quantum Chemistry package
\cite{MOLPRO}.

\begin{figure} [htbp]
{\includegraphics[angle=270,width=\columnwidth,clip]{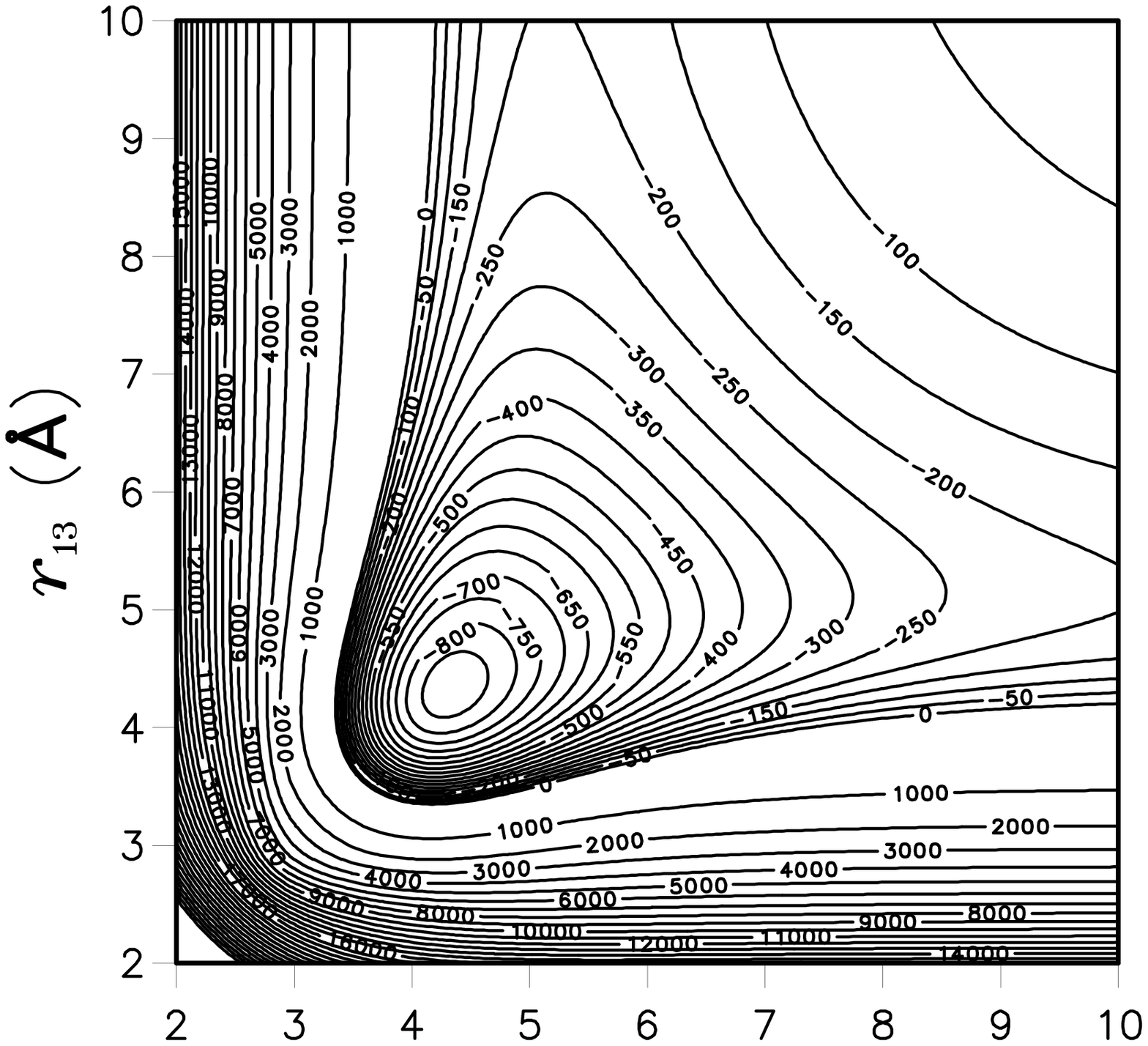}}
{\includegraphics[angle=270,width=\columnwidth,clip]{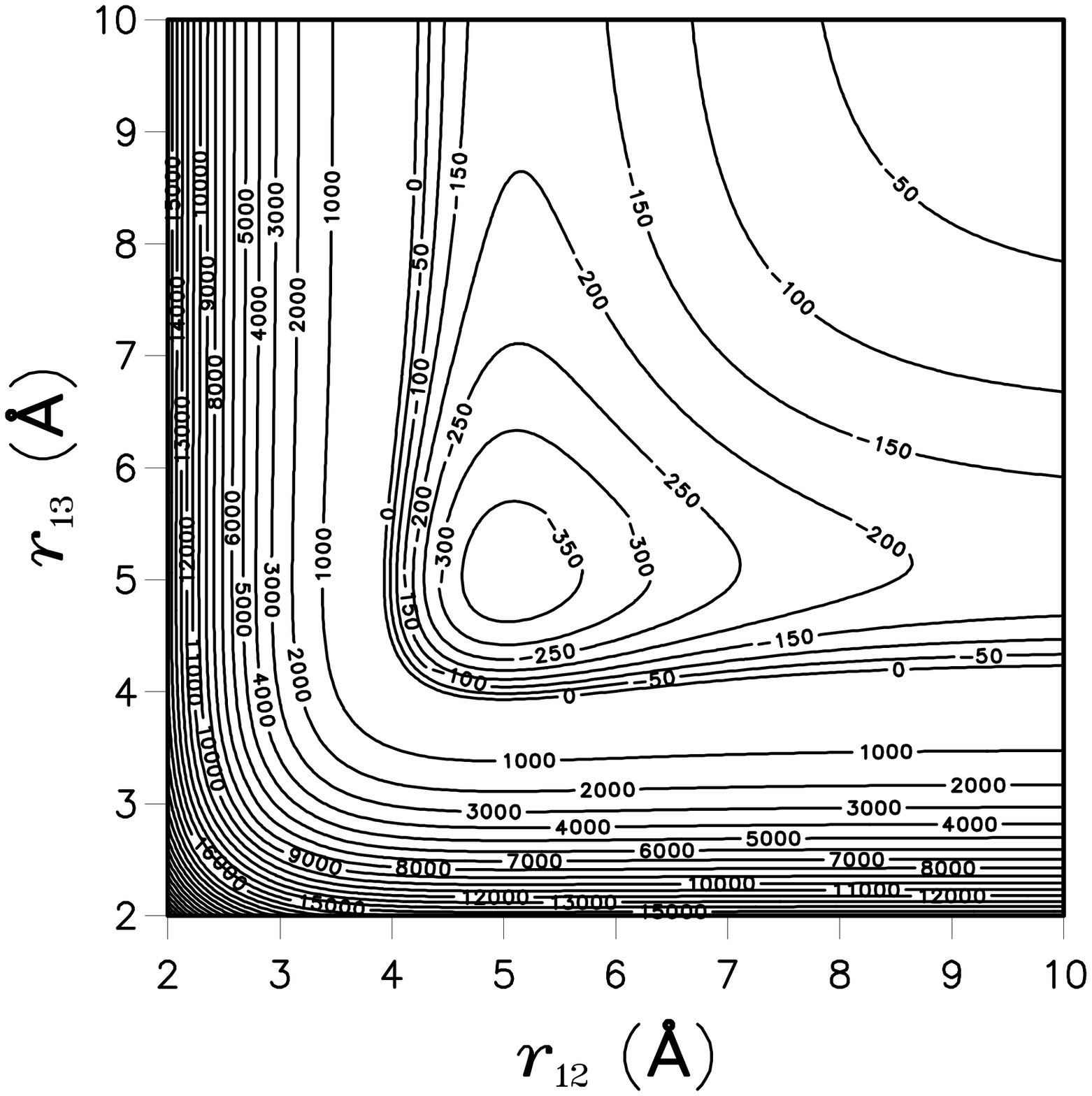}}
\caption{Cuts through the Na$_{3}$ quartet surface in valence
coordinates. Upper panel: cut for a bond angle of 60$^\circ$;
the global minimum of $-881$ cm$^{-1}$ is at $r_{12}=r_{13}=r_{23}=4.34$ \AA.
Lower panel: cut at collinear geometries; the collinear minimum of $-382$ cm$^{-1}$ is at $r_{12}=r_{13}=5.06$ \AA. Contours are labeled in cm$^{-1}$.} \label{na3pot}
\end{figure}

47 dimer interaction energies $V_{\rm dimer}$ on the $a^{3}\Sigma_{u}^{+}$
manifold were calculated on an irregular grid covering the range of
interatomic distances from 2.0\ \AA\ to 14.0\ \AA. These points were
interpolated using the 1D reciprocal-power reproducing kernel Hilbert
space (RP-RKHS) method~\cite{Ho96}. The interpolation was done with
respect to $r^{2}$ using RP-RKHS parameters $m=2$ and $n=3$). The
resulting curve had a minimum at approximately $r_{e} = 5.194$ \AA,
$V_{\rm dimer}(r_{e}) = - 172.946$ cm$^{-1}$, which is slightly higher
than the previously reported \textit{ab initio} minima $r_{e} = 5.192$
\AA, $V_{\rm dimer}(r_{e}) = - 177.7$ cm$^{-1}$~\cite{Gut99}, $r_{e}
= 5.20$ \AA, $V_{\rm dimer}(r_{e}) = - 176.17$ cm$^{-1}$~\cite{Hig00},
and $r_{e} = 5.214$ \AA, $V_{\rm dimer}(r_{e}) = - 174.025$ cm$^{-1}$
\cite{Sol03}.

Ivanov \textit{et al.}~\cite{Na2bench} analyzed experimental data
on triplet Na$_{2}$ and derived the accurate position $r_{e} =
5.16607$ \AA, $V_{\rm dimer}(r_{e}) = - 173.64960$ cm$^{-1}$ of
the $a^{3}\Sigma_{u}^{+}$ minimum.  Therefore our \textit{ab initio}
interaction energies were shifted and scaled (shifted by -0.02754\ \AA\
and scaled by 1.00407) so that the minimum of the modified potential
energy curve coincided with the minimum determined from experiment. The
RP-RKHS interpolation was then repeated using the modified RP-RKHS
method~\cite{Ho00}.  Beyond the last \textit{ab initio} point, the
potential energy was then extrapolated to the form
\begin{equation}
\label{Eq2} V_{\rm dimer}(r) = -\frac{C_{6}}{r^{6}} - \frac{C_{8}}{r^{8}} -
\frac{C_{10}}{r^{10}}.
\end{equation}
The long-range coefficients $C_{6}$ and $C_{8}$ were kept fixed to the
values of $1.561\times10^{3}$ $E_{h}\,a_{0}^{6}$ and $1.16\times10^{5}$
$E_{h}\,a_{0}^{8}$, respectively~\cite{Mit03}. The value of the
``free'' long-range coefficient $C_{10}$ was then determined from the
corresponding RP-RKHS coefficients~\cite{Sol00} to be $1.19\times10^{7}$
$E_{h}\,a_{0}^{10}$, which compares very well with $1.158\times10^{7}$
$E_{h}\,a_{0}^{10}$ from Ref.\ \cite{Mit03}. The resulting potential
energy curve supports 16 vibrational bound states and gives a scattering
length of 67.1\ $a_0$, which compares reasonably well (within 10\%)
with published values 65.3~\cite{Ver99}, 63.9~\cite{Jul00}, and 62.51
\cite{Tie00}.

356 trimer interaction energies $V_{\rm trimer}$ on the $1^{4}A_{2}'$
manifold were calculated on a regular 3D grid covering the range
of interatomic distances from 2.5\ \AA\ to 10.0\ \AA\ 
(geometry configurations  were unique up to a permutation of atoms and satisfied the triangular inequality
$|r_{12}-r_{13}|\leq r_{23} \leq r_{12}+r_{13}$;
at linear geometries, where $r_{23}$ =$r_{12}$+$r_{13}$, the distance $r_{23}$ was permitted to extend beyond 10.0\ \AA).

The grid
consisted of 220 C$_{\rm 2v}$ points (including 16 D$_{\rm 3h}$ points)
and 136 C$_{\infty{\rm v}}$ points (including 16 D$_{\infty{\rm h}}$
points). From the trimer interaction energies the counterpoise corrected
non-additive energies $V_{3}$ were extracted using Eq.\ (\ref{Eq1}).

The non-additive energy function $V_{3}$ was represented in the
same manner as in the case of the spin-polarized potassium trimer
\cite{Que05,Cvi06}.  In order to accommodate the geometric dependencies
of the long-range multipole terms, third-order dipole-dipole-dipole
\cite{ATM} and dipole-dipole-quadrupole~\cite{Bell} terms were subtracted
from the non-additive energy $V_{3}$. Their corresponding long-range
coefficients $C_{9}$ and $C_{11}$ were fixed to $1.892\times10^{5}$
$E_{h}\,a_{0}^{9}$~\cite{Mit03} and $1.46812\times10^{5}$
$E_{h}\,a_{0}^{11}$ respectively~\cite{PT97}. The leading term of
the remaining multipole asymptotic expansion was the fourth-order
dipole-dipole-dipole term~\cite{Bade}, and after a multiplication by a
suitable function it was prepared  for an ``isotropic'' extrapolation
\cite{Que05,Cvi06}.  The resulting points were then interpolated, using
the fully symmetrized 3D RP-RKHS interpolation method~\cite{Hig00},
in each interatomic distance with respect to the reduced coordinate
$\rho=\left(\frac{r}{S}\right)^{3}$ and with RP-RKHS parameters
$S=10.0$\,\AA,$\,m=0,\,n=2$.

The three-atom interaction potential $V_{\rm trimer}$ for the
$1^{4}A_{2}'$ state of Na$_{3}$ was then reconstructed using Eq.\
(\ref{Eq1}). Its D$_{3h}$ global minimum -880.9 cm$^{-1}$ is  at
$r_{12}=r_{13}=r_{23}=4.34$\ \AA\ and D$_{\infty h}$ saddle point -381.7
cm$^{-1}$ is at $r_{12}=r_{13}=5.06$\ \AA. The minimum of our trimer
potential is approximately 5\% deeper than the minima reported by Higgins
\textit{et al.}~\cite{Hig00} and Sold\'{a}n \textit{et al.}~\cite{Sol03}. Two cuts through the surface are
shown as contour plots in Fig.\ \ref{na3pot} for values of the
valence angle 60$^{\circ}$ and 180$^{\circ}$.

\section{Quantum dynamics}

The scattering observables are obtained by solving the time-independent
Schr{\"o}dinger equation for three atoms. Quantum dynamical calculation
are performed using hyperspherical democratic coordinates. This system of
coordinates comprises of three internal coordinates (two hyperangles and one
hyperradius) describing the shape and the size of the molecular triangle
and three Euler angles describing the orientation of the molecular plane
in space.  The total wavefunction is expanded on
a set of hyperspherical basis functions varying with the hyperradius.
The resulting closed coupled equations are solved using a log-derivative
propagator approach. Details on the method can be found in~\cite{jmlhyp}.
\begin{figure}[htbp!] \includegraphics[width=\columnwidth,clip]{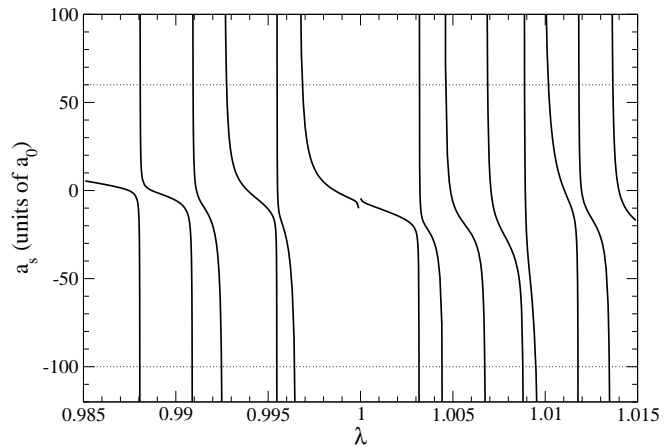}
\caption{
The Na+Na$_2(v=0,j=0)$ $s$-wave scattering length as a function of the
three-body control parameter $\lambda$ (see text). 
Horizontal lines identify
model potentials corresponding to $a_s=-100a_0$ and $60a_0$.
}
\label{fig1}
\end{figure}

The hyperspherical democratic coordinates are especially well adapted in describing
alkali species reactions that mainly proceed through an insertion
mechanism~\cite{Sol02}. However, the region of large interparticle
distances, where the system separates into atom and molecule, is
not efficiently described in hyperspherical coordinates. Therefore
the scattering wavefunction in the outer region is computed using
Jacobi coordinates. State-to-state probability amplitudes are finally
extracted by matching to the short-range wavefunction obtained with the
hyperspherical approach~\cite{jmlhyp}.

Atom-atom collisions in the ultracold regime are determined by a
small number of parameters. In fact, the $s$-wave scattering lengths and
the dominant term in the long-range multipole potential expansion
are sufficient in general to predict all near-threshold scattering
and bound state properties. This feature has allowed many systems
of experimental interest to be accurately modeled based on a limited
amount of experimental information~\cite{cs2,rb2,krb}.
Observed energy-dependent cross sections and Feshbach or shape resonances often provided a
key piece of information for determining theoretically the scattering
lengths and the long-range dispersion coefficients.

The situation appears to be more complex for atom-molecule collisions due
to the additional ro-vibrational degrees of freedom and anisotropic interactions.
As collisional data may be soon available we begin to study here
what experimental information might be best suited to constrain
the collision models. For most alkali systems, and for Na$_2$
in particular, the two body potential is known with high accuracy
from a combination of conventional room temperature and ultracold atom
spectroscopy~\cite{Tie00}. Therefore, one may expect the three-body
interaction $V_3$ to represent the largest source of uncertainty. We
assume that its shape is essentially correct and following the approach
of~\cite{Que04}, we solve the scattering problem with a scaled potential
$\lambda\,V_{3}$. At variance with Ref.~\cite{Que04} which considered inelastic
scattering we discuss here scattering resonances in elastic collisions.

We focus on molecules in the lowest triplet rovibrational state, 
which is collisionally stable under two-body collisions with atoms if
both colliding partners have maximal spin projection on the quantization axis.
We show in Fig.~\ref{fig1} the atom-diatom $s$-wave
scattering length $a_s$ as a function of the three-body control
parameter $\lambda$. Each time a three-body bound state crosses the
dissociation threshold the $a_{s}$ presents a typical divergence,
termed a zero-energy resonance. One may note that a 0.1-1\% potential variation 
(1-10~cm$^{-1}$ on the potential depth) is sufficient for a complete
$- \infty$ to $+ \infty$ variation of $a_s$. 

In order to investigate the relation between the zero-energy 
quantity $a_s$ and finite-energy scattering, we select values of the control
parameter $\lambda$ corresponding to the same value of $a_s$ and
compare the corresponding energy-dependent elastic cross sections.

\begin{figure}[htbp] \includegraphics[width=\columnwidth,clip]{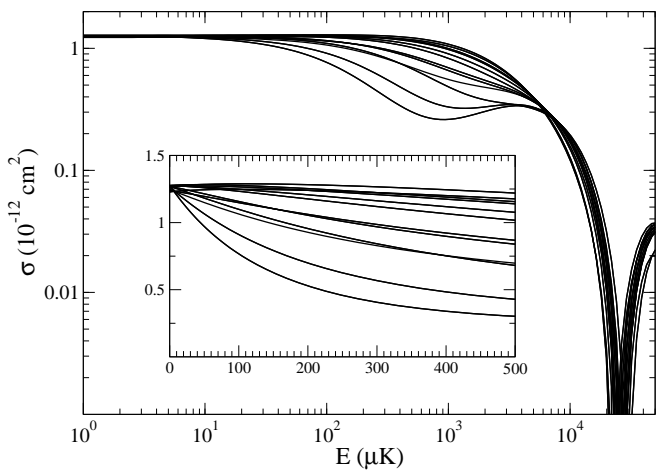}
\caption{
The partial $J=0$ Na+Na$_2(v=0,j=0)$ elastic cross section as a function
of collision energy for different values of the three-body control parameter determined by the
same value of the atom-diatom scattering length $a_s=60 a_0$. The inset refers to the typical energy
regime of current ultracold molecule experiments.
}
\label{fig2}
\end{figure}

We first consider total angular momentum $J=0$. For $j=0$ rotational states this
implies an angular momentum $\ell=0$.
Fig.~\ref{fig2} shows the result of the comparison for a typical positive
value of $a_s$. 
The
partial $J=0$ cross sections $\sigma$ for $a_s>0$ show a qualitatively similar
behavior essentially determined by the value of $a_s$ and by the
long-range $C_6$ coefficient. 
One can remark
the well known zero energy limit $\sigma \to 4\pi a_s^2$. 
The minimum of the cross
section corresponds to the scattering phase shift going through a multiple of
$\pi$, and in the absence of contributions from higher order
partial waves would correspond to a Ramsauer-Townsend minimum in the total cross section~\cite{Mott}.

The cross sections calculated for $a_s<0$ are more interesting. A large negative
scattering length is associated with a bound state turned into a virtual
state with energy $E_0$ in the continuum. This situation can be conveniently described
by decomposing the elastic phase shift
into a background plus a resonance contribution \cite{Mott,Mies00}
\be 
\delta= \delta_{\rm bg} + \delta_{\rm res}
  \quad,\quad \delta_{\rm res}=-\arctan{ \frac{\gamma/2}{E-E_{r} } }
\label{tres}
 \ee
where $E_r=E_0+\eta$ is the resonance position, with $\gamma$ and $\eta$ the resonance width
and shift, respectively.
For the first few angular momenta $\ell$ low energy scattering is determined by the asymptotic  
behavior $ \gamma \sim E^{\ell + 1/2}$,
$\delta_{\rm bg} \sim E^{\ell + 1/2}$ and $\eta \sim {\rm const}$ \cite{Mies00}.

There will be a low-energy resonance only if the width is sufficiently
narrow $\gamma \left( E \right) < E$. If the more strict condition $\gamma \left( E \right) \ll E$ 
is fulfilled $\gamma\left(
E \right)$ can be replaced with the constant quantity $\gamma_r = \gamma\left( E_r
\right)$ and the decomposition in Eq. (\ref{tres}) implies that $\delta$
undergoes a rapid $\pi$ variation across resonance.
Note that because of the
$\gamma \sim E^{1/2}$ threshold law, $\ell=0$
scattering at sufficiently low energy always violates the $\gamma < E$ 
condition and no resonant behavior will arise. 
However, this does not rule out the presence of 
resonances at higher yet very low collision energy (see below).

Resonances can also be analyzed in terms of the Wigner
time delay~\cite{wigner}, \be Q=2 \hbar
\frac{d \delta}{dE}, \ee {\it i.e.} the average delay of a scattering event compared to
free transit in the absence of the potential.
\begin{figure}[htbp] \includegraphics[width=\columnwidth,clip]{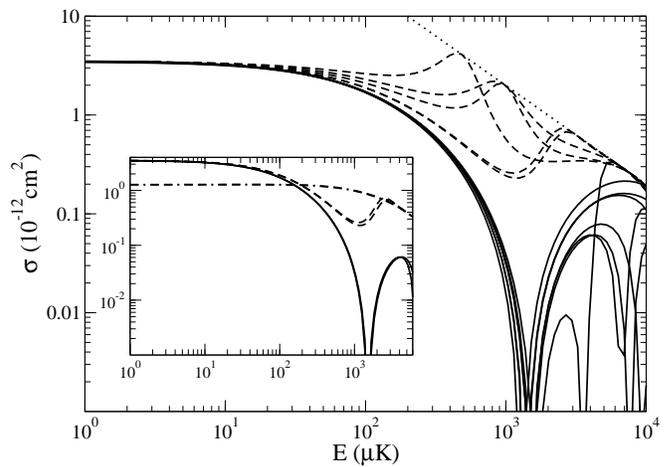}
\caption{
The partial $J=0$ Na+Na$_2(v=0,j=0)$ elastic cross section as a function
of collision energy for different values of the three-body control parameter determined by the
same value of the atom-diatom scattering length $a_s=-100 a_0$. Cross sections presenting resonant behavior are
identified by dashed lines. The dotted line denotes the unitarity limit (see text).
Sample cross sections obtained using different $\lambda$ and presenting very similar energy 
dependence are emphasized in the inset. 
The inset also shows (dash-dotted lines) two virtually identical cross sections for $a_s>0$ extracted from Fig.~\ref{fig2}. 
}
\label{fig3}
\end{figure}
In the threshold regime using Eq.~(\ref{tres}) 
one obtains:
\bea 
Q&=&\frac{\hbar \gamma}{\left(E-E_r \right)^2 +
\frac{\gamma^2}{4}} \left( 1-\frac{d\eta}{dE} \right) 
+ \frac{2}{ v} \frac{d \delta_{\rm bg}}{dk}    \nonumber \\
&-& \frac{E-E_r}{(E-E_r)^2 + \frac{\gamma^2}{4}  } 
 \frac{1}{v } \frac{d \gamma}{dk} 
\label{eq5} 
\eea 
where $v$ is the
velocity in the relative motion and $k$ the relative wave vector. 
The first term is the usual
Lorentzian profile arising from exponential decay with a $\frac{d\eta}{dE}$ correction. 
Near isolated resonances at high energy this is usually the dominant contribution
to the time delay; see {\it e.g.}~\cite{aquila}. The resonance shift $\eta$ is usually 
a slowly varying function
of energy but in the presence of additional scattering features such as
shape resonances in the background continuum~\cite{2002-AS-PRA-063406}.

The second term in Eq.~(\ref{eq5}) is the classical time for the
relative particle to span a distance $2 \frac{d \delta_{\rm bg}}{dk}  $.
For $\ell =0$ elastic scattering $\delta_{\rm bg} \sim -k a_{\rm bg}$, and this term reduces to 
$ -2 a_{\rm bg}/v  $ corresponding to an attractive (repulsive) character 
of the background potential for negative (positive) background scattering lengths $a_{\rm bg}$.
The third term vanishes for $E=E_r$ and gives a correction of dispersive
shape across resonance. 

Note that in Fig.~\ref{fig3} one can essentially identify two classes of
curves. The first class (full lines) presents a monotonically decreasing behavior towards the first
minimum, which corresponds to nonresonant scattering. 
The second class (dashed lines) present 
peaks at which the scattering partial cross section reaches the unitarity limit $\sigma = \frac{4 \pi}{k^2} $ (dotted line in Fig.~\ref{fig2}) 
and can in principle be associated with a resonant behavior.
\begin{figure}[htbp] \includegraphics[width=\columnwidth,clip]{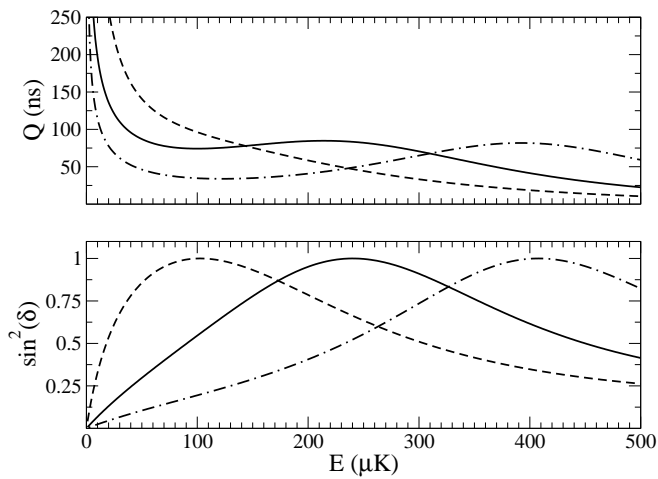}
\caption{
The Wigner time delay $Q$ (upper panel) and the $\sin^2 \delta$
quantity for $J=0$ Na+Na$_2(v=0,j=0)$
elastic collisions and selected $\lambda$ values. Only the peak near $400 \mu$K can
be classified as a resonance (see text).
}
\label{qten}
\end{figure}

We focus on the lowest-energy peak of Fig.~\ref{fig3} and make the three-body potential slightly
more attractive in order to further shift this feature towards threshold.
Fig.~\ref{qten} shows the $\sin^2  \delta  $ and $Q$ quantities for three selected potentials. The feature
near $400\mu$K can be essentially classified as a resonance with $\gamma_r
\simeq 0.5 E_r$. As the potential becomes more binding $\gamma $ becomes
larger than $E_r$ ($\gamma_r = E_r$ for $E_r\simeq 200 \mu$K) and the resonant behavior tends to disappear. As the
peak is made to shift closer to threshold the time delay
is fully dominated by the background contribution; see upper panel.
Also note (lower panel) that in all cases the unitarity limit $\sin^2  \delta  =1$ is attained.

An additional interesting feature that can be observed by inspection of Figs.~\ref{fig2}-\ref{fig3} is the
near coincidence of $J=0$ elastic cross sections computed with different
three-body potentials in the whole energy range $E<10$~mK. 
Sample cross sections illustrating this circumstance for both $a_s>0$ and $a_s<0$ are put forth in
the inset of Fig.~\ref{fig3}.
This shows that knowledge of energy dependent cross sections in the regime where only the $J=0$
partial wave is important is in general not sufficient to determine
the strength of the three-body nonadditive potential, even if its shape
were precisely fixed.

\begin{figure}[htbp] \includegraphics[width=\columnwidth,clip]{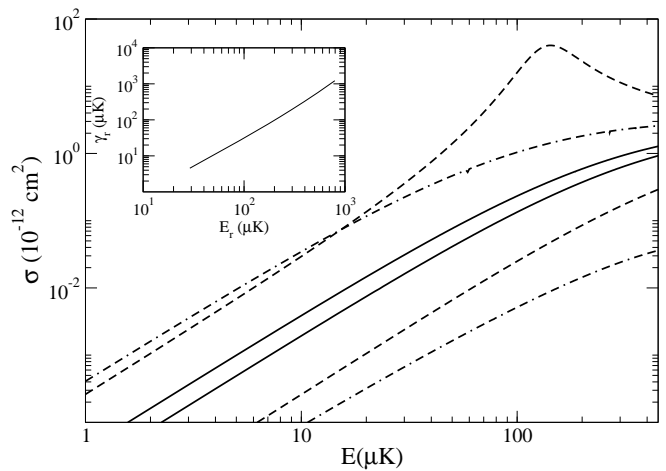}
\caption{
The partial $J=1$ Na+Na$_2(v=0,j=0)$ elastic cross section
calculated with the same set of model potentials used in the inset of Fig.~\ref{fig3}.
Same line style is used for corresponding cross sections obtained with the same potential.
A resonance can be observed in the upper dashed curve. The inset
shows the resonance width $\gamma_r$ as a function of its position $E_r$.
}
\label{fig3J1}
\end{figure}

However, model potentials giving equivalent $J=0$
cross sections will not in general be equivalent if $J>0$ scattering is
explored. For instance Fig.~\ref{fig3J1} shows the $J=1$ elastic 
cross sections for $(v=0,j=0)$ molecules. Calculations use the same model potentials as the inset 
in Fig.~\ref{fig3} and can
be identified according to the line style. Order of magnitude differences are observed for $J=1$ cross
sections in cases where $J=0$ cross sections are identical.
One can conclude that the initial characterization of a theoretical model based on purely elastic 
collisions should take into account a sufficiently broad energy range for the contribution of 
$J>0$ partial waves to become observable.
In alternative, in cases where the method for the production of cold molecules allows the initial 
ro-vibrational state to be controlled, at least one experimental inelastic cross section 
(for some $(v=0, j>0)$ initial state, for instance) should 
complement the elastic collision data.

Please note that one peak is also observed in Fig.~\ref{fig3J1}.
Analysis of numerical results based on Eq.~(\ref{tres}) shows that $\gamma_r \simeq 0.5 E_r$, {\it i.e.} this feature
represents a resonance.
The inset shows the resonance width as the peak center is shifted by making the $\lambda$ parameter vary.
Its position below the maximum $\simeq 400 \mu$K  of the
$\ell=1$ centrifugal barrier (we find $\gamma_r = E_r$ for $E_r \simeq 500 \mu$K) suggests that 
it is a shape resonance. However, Feshbach coupling similar to the one found for $J=0$ collisions 
is not conclusively ruled out.

In conclusion, we have presented a new potential energy surface for Na$_{3}(1^{4}A_{2}')$.
We have demonstrated that long lived triatomic complexes exist and give rise to resonance effects
in reactive collisions even at very low collision energies. 
General features to be expected
in atom-molecule scattering in the ultracold regime have also investigated by performing 
a systematic variation of the three-body part of the interaction potential.
Knowledge of energy-dependent $J=0$  elastic cross sections
may not be sufficient to determine 
the strength of the nonadditive three-body
interaction. To this aim, at least
one additional $J>0$ elastic or inelastic cross section needs to be experimentally determined. 
In this work we have studied the sensitivity of scattering observables by introducing a global
scaling parameter of the three-body interaction. In perspective, as cold collision empirical data will become
available, it is likely that more complex 
parametrizations of the potential energy surface will need to be introduced in order to compare quantitatively theory and experiments.

\begin{acknowledgments}
We wish to thank A.~Viel for useful discussions.
The authors acknowledge support of Egide (PHC Barrande \# 13860UA) and of the Ministry
of Education, Youth and Sports of the Czech Republic (KONTAKT
project Barrande 2-07-3 and Research project no.\ 0021620835).
PS appreciates support of the European Science Foundation and the Czech
Science Foundation (EUROCORES programme EuroQUAM, project QuDipMol,
grant no.\ QUA/07/E007).
\end{acknowledgments}

\end{document}